\documentclass[aps,prl,twocolumn,groupedaddress]{revtex4}

\usepackage{graphicx} 
\usepackage{dcolumn}
\usepackage{bm}

\begin{document}
\title{Polaron-bipolaron  superconductivity and the intrinsic pairing symmetry in cuprate superconductors } 
\author{Guo-meng Zhao$^{1,a}$ } 
\affiliation{$^{1}$Department of Physics and Astronomy, 
California State University, Los Angeles, CA 90032, USA~\\}
\begin{abstract}

We have calculated the tunneling conductance of a superconductor-insulator-superconductor junction based on the polaron-bipolaron theory of superconductivity. The predicted incoherent hump features are in quantitative agreement with tunneling spectra of optimally doped Bi$_{2}$Sr$_{2}$CaCu$_{2}$O$_{8+y}$ and Bi$_{2}$Sr$_{2}$Ca$_{2}$Cu$_{3}$O$_{10+y}$.  We further show that angle-resolved photoemission spectra of underdoped cuprates are consistent with the Bose-Einstein condensation of inter-site bipolarons and that the superconducting gap symmetry is $d$-wave, which is determined by the anomalous kinetic
process rather than by the pairing interaction. In the overdoped cuprates (BCS-like superconductors), the superconducting gap symmetry is the same as the pairing symmetry, which is found to be extended $s$-wave with eight line nodes in hole-doped cuprates and nodeless $s$-wave in electron-doped cuprates.  The polaronic effect  significantly enhances the density of states at the Fermi level and the effective electron-phonon coupling constant for low-energy phonon modes, which is the key to the understanding of high-temperature superconductivity.

\end{abstract}
\maketitle

\begin{center}
{\bf I.  Introduction}
\end{center}

Despite the fact that  many experiments have supported polaronic/bipolaronic 
superconductivity,  the bipolaronic theory of superconductivity has not 
been generally accepted. There are several reasons for this. First,  much attention 
has been paid to bipolaronic theories based on strong short-range Holstein-type
electron-phonon interactions (EPI). These theories fail to predict a high
critical temperature because the bipolarons are too heavy and localization of   
the heavy bipolarons is inevitable. Secondly, there is strong experimental evidence 
for the Cooper pairing rather than the real-space pairing  in optimally and overdoped 
cuprates \cite{Zhaoreview2}. Thirdly, angle-resolved photoemission spectra (ARPES), optical, and transport experiments have consistently 
pointed towards a large Fermi surface in overdoped cuprates \cite{Zhaoreview2}.    These results appear 
to argue against the bipolronic theory of superconductivity. However, an analytical
multi-polaron model of high-temperature superconductivity
in highly polarizable ionic lattices has recently been 
proposed with generic (bare) Coulomb and Fr\"ohlich EPI avoiding any ad-hoc 
assumptions on their
range and relative magnitude \cite{Alex2011}. The generic Hamiltonian comprising any-range 
Coulomb repulsion and the Fr\"ohlich EPI can be reduced to a short-range
$t-J_p$ model at very large lattice dielectric constant,
$\epsilon_{0}\rightarrow
\infty$, for the moderate and strong EPI. In this limit
the bare static Coulomb repulsion and EPI negate each
other giving rise to a novel physics described by the polaronic
$t-J_p$ model with a short-range polaronic spin exchange
$J_p$ of phononic origin \cite{Alex2011}. Moreover, if one considers realistic 
finite $\epsilon_{0}$, the cancelation of the bare Coulomb repulsion by the
Fr\"ohlich EPI is not complete so that a residual on-site repulsion $\tilde{U}$ of 
polarons could be
substantial if the size of the Wannier (atomic) orbitals
is small enough \cite{Alex2012}. The residual Hubbard $\tilde{U}$ drives the system to 
undergo a crossover from Bose-Einstein condensation (BEC) to the
Cooper pairing \cite{Alex2012}. This naturally reconciles the polaron-bipolaron theory of
superconductivity with the Cooper pairing and the large Fermi
surface observed in optimally doped and overdoped cuprate superconductors
\cite{Zhaoreview2}. The strong coupling of polarons with low-energy ($<$ 40 meV) multiple phonon modes is the key to  the understanding of high-$T_{c}$ mechanism in optimally doped cuprates \cite{Zhaoisotope}. It is interesting that similar polaronic mechanism has been applied to a nonmagnetic high-$T_{c}$ (30 K) superconductor Ba$_{1-x}$K$_{x}$BiO$_{3}$ (Ref.~\cite{ZhaoBKBO}). 

In the main text \cite{Maintext}, we have shown that  tunneling, ARPES, and neutron data cannot be consistently explained by magnetic pairing mechanism based on the spin-fermion model and on the $t-J$ model. This is in agreement with 
recent analytical \cite{AlexPRL} and numerical (Monte-Carlo) \cite{Aimi,Hardy} studies that cast serious doubts 
on the possibility of high temperature
superconductivity from repulsive interactions only. This is also in accord with optical 
experiments
\cite{YLee} which show that the
electron-boson spectral function 
$\alpha{^2}(\omega)$$F(\omega)$ is independent of magnetic field, in
contradiction with the theoretical prediction based on the magnetic
pairing mechanism (see Fig.~9 of Ref.~\cite{YLee}).  

\begin{center}
{\bf II. Break-junction tunneling spectra calculated according to the polaron-bipolaron theory}
\end{center}

According to the polaron-bipolaron theory of superconductivity, the tunneling conductance $dI/dV$ of a superconductor-insulator-superconductor (SIS) junction is given by \cite{AlexEPL}
\begin{eqnarray}
dI/dV\propto &\int_{0}^{\infty}dt\exp[2g^{2}e^{-(\delta\Omega) t}\cos(\Omega t)-\Gamma t]\nonumber \\
&\cos[2g^{2}e^{-(\delta\Omega) t}\sin(\Omega t)-(e|V|-2\Delta)t],
\end{eqnarray}
where $\Omega$ is the characteristic energy of phonon modes in  the polaron cloud, $\delta\Omega$ is the phonon dispersion, $g^{2}$ is the average number of phonons in the polaron cloud, which is proportional to the electron-phonon coupling strength, $\Gamma$ is the life-time broadening parameter, $\Delta$ = $\sqrt{\Delta^{2}_{p} + \Delta^{2}_{c}}$, $\Delta_{p}$ is the single-particle gap (normal-state gap), and $\Delta_{c}$ is the superconducting gap that closes at $T_{c}$. The tunneling conductance can be readily calculated using numerical integration.

\begin{figure}[htb]
    \includegraphics[height=12cm]{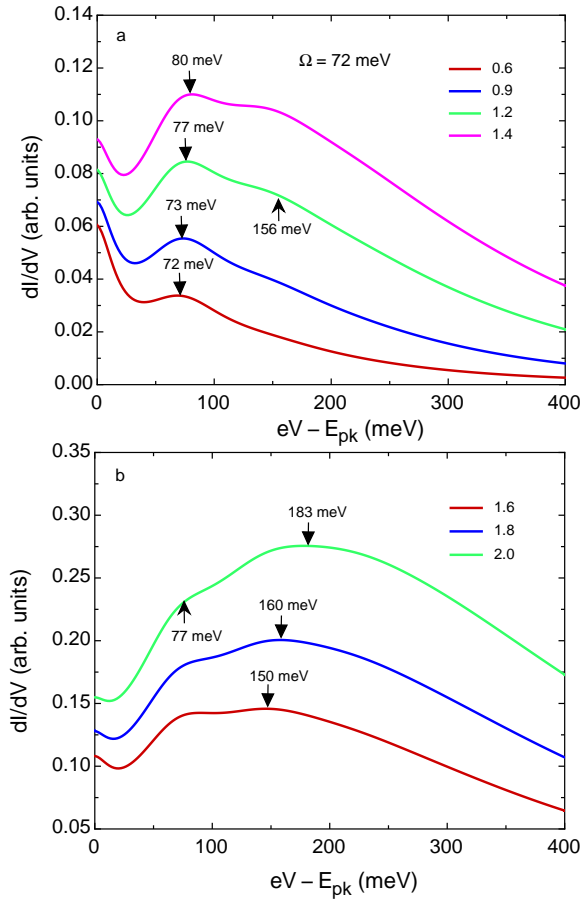}
 \caption[~]{Numerically calculated SIS tunneling spectra  with varying $g^{2}$. The values of $g^{2}$ are labeled by the numbers in the figure. The energy of the coherence peak position ($E_{pk}$) is close to 2$\Delta$. The down-arrows mark the dominated hump features.  }
\end{figure}

Figure 1 shows the numerically calculated SIS tunneling spectra with varying $g^{2}$ and fixed $\Omega$ = 72 meV, $\delta\Omega$  = 30 meV, $\Delta$ = 35 meV, and $\Gamma$ = 20 meV. The energy of the coherence peak position ($E_{pk}$) is slightly higher than 2$\Delta$ due to a finite $\Gamma$ value. As $g^{2}$  increases, the spectral weight of the coherence peak  is reduced while the weight of the  incoherent broad hump feature increases. The simulated coherence peak  is much lower than the experimental one \cite{Dewi} due to the fact that our calculation assumes a constant density of states at the Fermi level, in contrast to the existence of the extended van Hove singularity proximity to the Fermi level. Another interesting result is that double hump features occur at about $E_{pk} +\Omega$ and $E_{pk} + 2\Omega$, respectively.  For $g^{2}$ $<$ 1.5, the hump feature at about $E_{pk} +\Omega$ is dominated while for $g^{2}$ $>$ 1.6 the hump feature at about $E_{pk} + 2\Omega$ is dominated. Fig.~6a in the main text \cite{Maintext} shows the energy separation between the dominated hump feature and the coherence peak as a function of $g^{2}$. 

\begin{figure}[htb]
    \includegraphics[height=12cm]{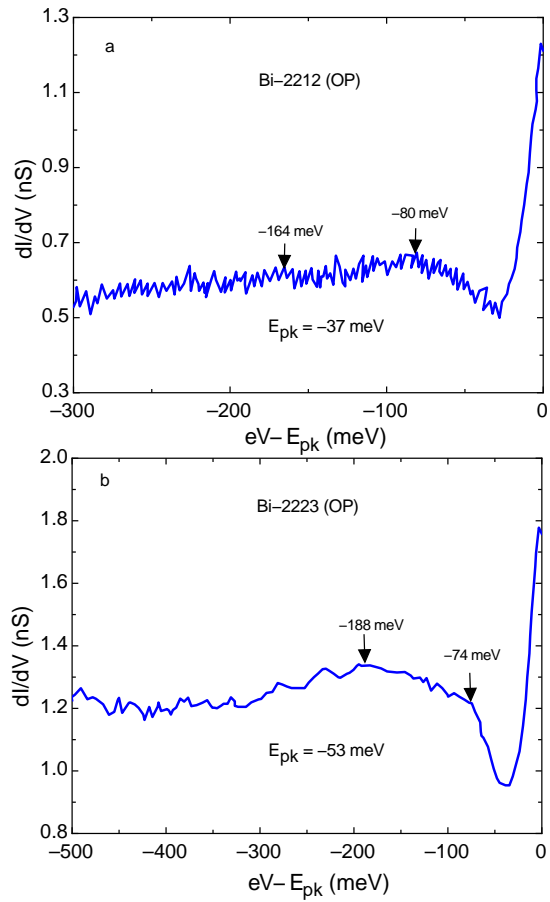}
 \caption[~]{Scanning  tunneling spectra  in the negative bias for optimally doped Bi$_{2}$Sr$_{2}$CaCu$_{2}$O$_{8+y}$ (Bi-2212) and Bi$_{2}$Sr$_{2}$Ca$_{2}$Cu$_{3}$O$_{10+y}$ (Bi-2223). The spectra were reproduced from Refs.~\cite{Dewi,Kug}. }
\end{figure}

It was shown that scanning tunneling spectra in the negative bias  are very similar to  break-junction tunneling spectra except that the coherence peak position in scanning tunneling spectra is at about $\Delta$ while it is at about 2$\Delta$ in break-junction tunneling spectra \cite{Dewi}. It was also found that the scanning tunneling spectra in the negative bias are nearly identical to the photoemission energy distribution curves (EDCs)  along the antinodal direction \cite{Ding2}. Therefore, in scanning tunneling spectra and antinodal EDCs, the two hump features should occur at about $E_{pk} +\Omega$ and $E_{pk} + 2\Omega$, respectively. Due to a low energy-resolution in angle-resolved photoemission 
spectra, one could see a local maximum at about $E_{pk} +\Omega$, at about $E_{pk} + 2\Omega$, or at about $E_{pk} + 1.5\Omega$ (the midway of two equal-hight hump features). In contrast, one should be able to see double hump features in scanning tunneling spectra due to a high energy-resolution. 

In Figure 2, we reproduce scanning  tunneling spectra  in the negative bias for optimally doped Bi$_{2}$Sr$_{2}$CaCu$_{2}$O$_{8+y}$ (Bi-2212) and Bi$_{2}$Sr$_{2}$Ca$_{2}$Cu$_{3}$O$_{10+y}$ (Bi-2223). It is apparent that both spectra show double hump features, as indicated by the arrows. For optimally doped Bi-2212, the hump feature at about $E_{pk} +\Omega$ is dominated, suggesting that $g^{2}$ is about 1.2. For optimally doped Bi-2223, the hump feature at about $E_{pk} + 2\Omega$ is dominated, suggesting that $g^{2}$ is about 2.0. It appears that the $E_{pk}$ value is approximately proportional to the $g^{2}$ value, in agreement with the polaron-bipolaron theory of superconductivity.  Furthermore, the energy positions of the double hump features in the tunneling spectra of the two optimally cuprates are in quantitative agreement with the theoretical predictions (see the green lines in Fig.~1a and Fig.~1b). We do not believe that any other theoretical model can explain the tunneling spectra.

\begin{center}
{\bf III. Intrinsic pairing  symmetry }
\end{center}

Another important issue is the intrinsic pairing gap symmetry
in the bulk of superconducting cuprates, which is directly related to the pairing interaction and pairing mechanism.  The symmetry of the superconducting condensate (gap), which  can be directly probed by phase-sensitive experiments such as Josephson-junction experiments, is not necessarily the same as the pairing gap symmetry.   Within the Bardeen-Cooper-Schrieffer (BCS) picture, both  symmetries happen to be the same. But in the strong-coupling limit, real-space pairing becomes possible and the symmetry of the superconducting condensate is completely different from the pairing gap symmetry \cite{Alex98,Wei}. Without differentiating between the pairing gap symmetry and the symmetry of the superconducting condensate, the $d$-wave symmetry of the superconducting condensate probed by the in-plane phase-sensitive experiments for both electron- and hole-doped cuprates \cite{Tsuei} has mistakenly been taken as indisputable evidence for  $d$-wave magnetic pairing mechanism.

\begin{figure}[htb]
    \includegraphics[height=7cm]{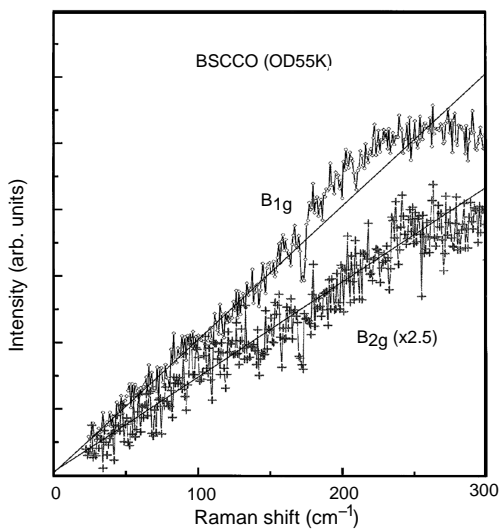}
 \caption[~]{Raman spectra in the $B_{1g}$ and $B_{2g}$ symmetries in an overdoped BSCCO with $T_{c}$ = 55 K. The figure is reproduced from Ref.~\cite{Hew}. }
\end{figure}

On the other hand, based on the quantitative analyses of many bulk-sensitive
experiments in optimally doped and overdoped cuprates, we
have concluded that the intrinsic pairing gap symmetry in the bulk
of cuprates is not $d$-wave, but extended $s$-wave (having eight
line nodes) in hole-doped cuprates \cite{Zhao01,ZhaoPS} and nodeless s-wave in electron-doped cuprates \cite{Zhao01,Zhao10,ZhaoJP}. In particular, a bulk and phase-sensitive experiment in an overdoped Bi$_{2}$Sr$_{2}$CaCu$_{2}$O$_{8+y}$ (BSCCO) along $c$-axis supports $s$-wave gap symmetry \cite{Li}. In fact, this phase-sensitive experiment is in quantitative agreement with an extended $s$-wave gap \cite{Zhao01}. Bulk-sensitive Raman spectra in the $B_{1g}$ and $B_{2g}$ symmetries for a heavily overdoped BSCCO with $T_{c}$ = 55 K (Fig.~2) provides indisputable evidence for an extended $s$-wave gap: $\Delta$ = 15$\cos 4\theta$~meV (where $\theta$ is the angle measured from the Cu-O bonding direction). The linear energy dependence of both $B_{1g}$ and $B_{2g}$ spectra is neither consistent with an isotropic $s$-wave gap nor with a $d$-wave gap.  Earlier angle-resolved photoemission spectra  for a heavily overdoped BSCCO with $T_{c}$ = 60 K also suggested a large gap $\Delta_{D}$ (about 10 meV) along the diagonal direction \cite{Vob}. On the other hand,  ARPES data for a slightly overdoped BSCCO \cite{Ding} are consistent with either $d$-wave or an extended $s$-wave gap with  $\Delta_{D}$ = 4.0$\pm$2.5 meV. More recent high-resolution ARPES data for slightly overdoped Bi$_{2}$Sr$_{2}$CaCu$_{2}$O$_{8+y}$  
and Bi$_{2}$Sr$_{2}$Ca$_{2}$Cu$_{3}$O$_{10+y}$  confirm finite gap sizes along the diagonal direction (see Fig.~4). The symmetrized  EDCs shown in Fig.~4 clearly show double peak features, which definitely proves a finite gap along the diagonal direction in both compounds. Quantitative  analyses of the EDCs \cite{ZhaoPS} show that  $\Delta_{D}$ is 6.0 meV for slightly overdoped Bi-2212, 14.0 meV for a heavily overdoped BSCCO with $T_{c}$ = 60 K, and 7.0 meV for slightly overdoped Bi-2223. Therefore, the pairing gap symmetry in overdoped double-layer and three-layer cuprates is extended $s$-wave with eight line nodes. It is worth noting that disorder and/or contamination on the cleaved surface can readily suppress the gap to zero in  an extended angle-range centered at $\theta$ =~45$^{\circ}$.

\begin{figure}[htb]
    \includegraphics[height=12cm]{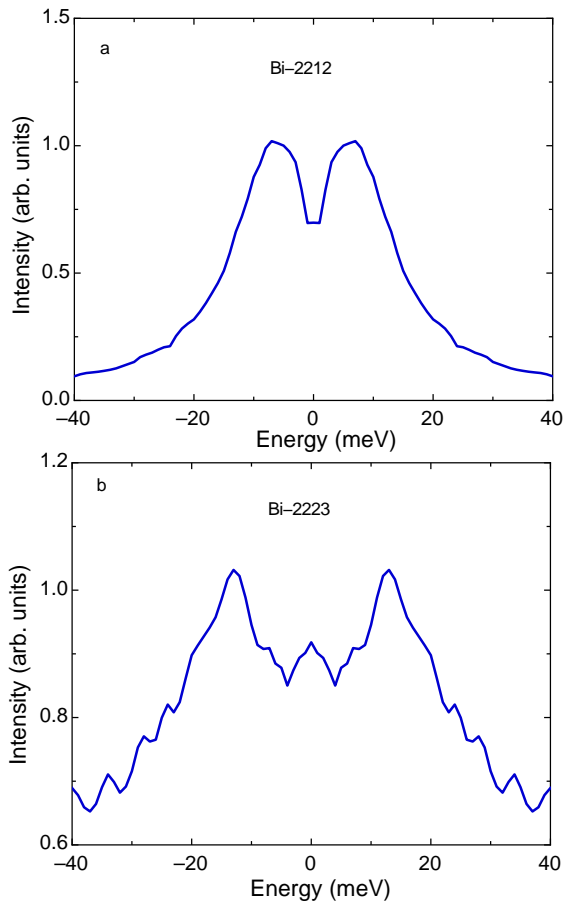}
 \caption[~]{The symmetrized  energy distribution curves (EDCs) along the diagonal direction for slightly overdoped Bi-2212 and Bi-2223. The data are taken from Refs.~\cite{Kor,Feng}. }
\end{figure}

On the other hand, EDCs for underdoped double-layer BSCCO show normal-state pseudogaps in the antinodal region and zero gap in an extended angle-range centered at $\theta$ = 45$^{\circ}$ (Ref.~\cite{Shennature}). In the superconducting state,  zero gap is seen at $\theta$ = 45$^{\circ}$, consistent with a $d$-wave gap symmetry \cite{Shen10}.  It is interesting that the pseudogap state even exists in a slightly overdoped single-layer (Bi,Pb)$_{2}$(Sr,La)$_{2}$CuO$_{6+y}$ (Bi-2201) with $T_{c}$ = 29~K (Ref.~\cite{Kondo}). These ARPES data can be well explained in terms of a local-pair superconductivity model, where inter-site pairs (or inter-site bipolarons) are condensed into superfluid through the Bose-Einstein condensation. Within this model, the superconducting gap $\Delta_{c}$ has a $d$-wave symmetry, which is controlled by the anomalous kinetic
process rather than by the pairing interaction \cite{Wei}. Because $\Delta_{c}$ in the antinodal region is much smaller than the pseudogap $\Delta_{p}$ \cite{Wei}, the total gap $\Delta$ = $\sqrt{\Delta^{2}_{p} + \Delta^{2}_{c}}$ is slightly larger than the normal-state pseudogap in this angle region, in agreement with the data \cite{Shennature,Kondo}. In contrast, $\Delta_{p}$ is zero in an extended angle-range centered at $\theta$ = 45$^{\circ}$ (possibly due to the pair-breaking effect in the presence of  impurities and disorder), so the total gap $\Delta$ is simply equal to $\Delta_{c}$ in this angle region. The fact that $\Delta$ $\geq$ $\Delta_{p}$ in the antinodal region of underdoped cuprates  \cite{Shennature,Kondo} provides strong evidence that the normal-state gap is associated with the superconducting pairing rather than with charge-density wave.

The local-pair superconductivity model for underdoped cuprates can also naturally explain  the $d$-wave superconducting gap symmetry probed by surface- and phase-sensitive experiments
based on planar Josephson tunneling. Because the surface- and phase-sensitive experiments are probing the superconducting state at surfaces and interfaces which were found to be
intrinsically underdoped \cite{Mann,Bet}, they naturally see the $d$-wave superconducting gap symmetry in underdoped cuprates.   Since the superconducting gap symmetry in underdoped cuprates is controlled by the anomalous kinetic
process rather than the pairing interaction \cite{Wei}, these phase-sensitive experiments do not determine the
pairing symmetry associated with the pairing interaction. Only for overdoped cuprates (BCS-like superconductors), the superconducting gap symmetry is associated with the pairing symmetry, so the $s$-wave pairing gap symmetry identified for overdoped cuprates \cite{Zhao01,ZhaoPS,Zhao10,ZhaoJP} places an essential constraint on the pairing interaction and rules out $d$-wave magnetic pairing mechanism. 

\begin{center}
{\bf  IV. Pairing interactions}
\end{center}
Now we address a basic question concerning the
key pairing interactions in cuprates.  Some
density functional (DFT) calculations \cite{Cohen,Heid} found small
EPI, which is too small  to explain high critical temperatures while some other first-principles studies found large EPI
in cuprates \cite{Bauer}.  It is common  that DFT underestimates
the role of the Coulomb interactions.  The inclusion of a short-range
repulsion (Hubbard U) via the LDA+U algorithm
\cite{Zhang} and/or nonadiabatic effects \cite{Bauer} significantly enhances
the EPI strength due to a poor screening of some particular
phonons. Furthermore,  due to layered structures, electron-phonon coupling to $c$-axis phonons cbecomes significant due to the
modulated long-range Madelung potential. The detailed discussion on the strong electron-phonon coupling to $c$-axis phonons can be found in a review article \cite{Johnston}. What is more important is the polaronic effect, which significantly enhances the density of states at the Fermi level and the effective electron-phonon coupling constant for low-energy phonon modes. That is the key to the understanding of high-temperature superconductivity \cite{Zhaoisotope}.

~\\
~\\
$^{a}$~gzhao2@calstatela.edu~\\

\bibliographystyle{prsty}

\end{document}